\documentclass[letterpaper, 10 pt, conference]{ieeeconf}  

\IEEEoverridecommandlockouts                              
\usepackage[T1]{fontenc}
\usepackage{graphicx}      
\usepackage{amsmath}
\usepackage{amssymb}
\usepackage{mathtools}
\usepackage{mleftright}
\usepackage[utf8]{inputenc}
\usepackage{prettyref}
\usepackage{xcolor}
\usepackage[permil]{overpic}
\usepackage{booktabs}
\usepackage{multirow}
\usepackage{ifthen}
\usepackage{xspace}
\usepackage{siunitx}
\usepackage{nicefrac}
\usepackage{tikz}
\usepackage{pgfplots}
\usetikzlibrary{external}
\usetikzlibrary{pgfplots.groupplots}
\allowdisplaybreaks

\newtheorem{rem}{Remark}

\newcommand{\overbar}[1]{\mkern 1.5mu\overline{\mkern-1.5mu#1\mkern-1.5mu}\mkern 1.5mu}

\newcommand{\dd}{\mathrm{d}}

\newcommand{\minimize}[2]{\underset{#2}{\mathrm{min}} #1}

\DeclareRobustCommand{\vec}[1]{ 				
	\ifthenelse{\equal{#1}{\omega} \OR \equal{#1}{\varphi} \OR \equal{#1}{\alpha} \OR \equal{#1}{\beta} \OR \equal{#1}{\chi} \OR \equal{#1}{\delta} \OR \equal{#1}{\varepsilon} \OR \equal{#1}{\phi} \OR \equal{#1}{\epsilon} \OR \equal{#1}{\gamma} \OR \equal{#1}{\eta} \OR \equal{#1}{\iota} \OR \equal{#1}{\kappa} \OR \equal{#1}{\lambda} \OR \equal{#1}{\mu} \OR \equal{#1}{\nu} \OR \equal{#1}{\pi} \OR \equal{#1}{\theta} \OR \equal{#1}{\vartheta} \OR \equal{#1}{\rho} \OR \equal{#1}{\sigma} \OR \equal{#1}{\varsigma} \OR \equal{#1}{\tau} \OR \equal{#1}{\upsilon} \OR \equal{#1}{\xi} \OR \equal{#1}{\psi} \OR \equal{#1}{\zeta} \OR \equal{#1}{\mathcal{A}}  \OR \equal{#1}{\mathcal{B}} \OR \equal{#1}{\mathcal{C}} \OR \equal{#1}{\mathcal{D}} \OR \equal{#1}{\mathcal{E}} \OR \equal{#1}{\mathcal{F}} \OR \equal{#1}{\mathcal{G}} \OR \equal{#1}{\mathcal{H}} \OR \equal{#1}{\mathcal{I}} \OR \equal{#1}{\mathcal{J}} \OR \equal{#1}{\mathcal{K}} \OR \equal{#1}{\mathcal{L}} \OR \equal{#1}{\mathcal{M}} \OR \equal{#1}{\mathcal{N}} \OR \equal{#1}{\mathcal{O}}
 \OR \equal{#1}{\mathcal{P}} \OR \equal{#1}{\mathcal{Q}} \OR \equal{#1}{\mathcal{R}} \OR \equal{#1}{\mathcal{S}} \OR \equal{#1}{\mathcal{T}} \OR \equal{#1}{\mathcal{U}} \OR \equal{#1}{\mathcal{V}} \OR \equal{#1}{\mathcal{W}} \OR \equal{#1}{\mathcal{X}} \OR \equal{#1}{\mathcal{Y}} \OR \equal{#1}{\mathcal{Z}}}{
		\boldsymbol{#1}
	}{
		\mathbf{#1}
	}
}

\newrefformat{sec}{Section~\ref{#1}}
\newrefformat{fig}{Fig.~\ref{#1}}
\newrefformat{tab}{Tab.~\ref{#1}}
\newrefformat{the}{Theorem~\ref{#1}}
\newrefformat{lem}{Lemma~\ref{#1}}
\newrefformat{rem}{Remark~\ref{#1}}
\newrefformat{app}{Appendix~\ref{#1}}
\newrefformat{ass}{Assumption~\ref{#1}}
\newrefformat{pro}{Proposition~\ref{#1}}

\definecolor{acin_red}{RGB}{186, 18, 43}
\definecolor{acin_gray}{RGB}{176, 176, 176}
\definecolor{acin_yellow}{RGB}{252, 204, 71}
\definecolor{acin_green}{RGB}{0, 190, 65}
\definecolor{TU_blue}{RGB}{0, 102, 153}
\definecolor{test2}{rgb}{0,0.47,0.85}
\definecolor{TU_gray}{RGB}{102, 102, 102}
\providecolor{test}{rgb}{0.85, 0, 0.22}

\title{\LARGE \bf
Iterative shaping of optical potentials for\\one-dimensional Bose-Einstein condensates
}

\author{Andreas Deutschmann-Olek, Mohammadamin Tajik, Martino Calzavara,\\Jörg Schmiedmayer, Tommaso Calarco, and Andreas Kugi
\thanks{This work was not supported by any organization}
\thanks{A. Deutschmann-Olek and A. Kugi are with the Complex Dynamical Systems Group, Automation and Control Institute, TU Wien, Vienna, Austria {\tt\small \{deutschmannn,kugi\}@acin.tuwien.ac.at}}%
\thanks{M. Tajik and J. Schmiedmayer are with the Atomic Physics and Quantum Optics Group, Atominstitut, TU Wien, Vienna, Austria {\tt\small mohammadamin.tajik@tuwien.ac.at}, {\tt\small schmiedmayer@atomchip.org}}%
\thanks{M. Calzavara and T. Calarco are with the Quantum Control Division, Peter Grünberg Institut, Forschungszentrum Jülich, Jülich, Germany {\tt\small \{m.calzavara,t.calarco\}@fz-juelich.de}}%
}

\begin{document}

\maketitle
\thispagestyle{empty}
\pagestyle{empty}

\begin{abstract}
The ability to manipulate clouds of ultra-cold atoms is crucial for modern experiments on quantum many-body systems and quantum thermodynamics as well as future metrological applications of Bose-Einstein condensate. While optical manipulation offers almost arbitrary flexibility, the precise control of the resulting dipole potentials and the mitigation of unwanted disturbances is quite involved and only heuristic algorithms with rather slow convergence rates are available up to now. This paper thus suggests the application of iterative learning control (ILC) methods to generate fine-tuned effective potentials in the presence of uncertainties and external disturbances. Therefore, the given problem is reformulated to obtain a one-dimensional tracking problem by using a quasi-continuous input mapping which can be treated by established ILC methods. Finally, the performance of the proposed concept is illustrated in a simulation scenario.
\end{abstract}


\section{Introduction}
\label{sec:introduction}

%
%
%

Large ensembles of ultra-cold atoms that form Bose-Einstein condensates (BECs) are nowadays routinely produced in labs around the globe to explore their complex quantum many-body behavior.
A particularly versatile and robust platform for trapping and manipulation of ultra-cold gases are magnetic traps using atom chips, i.e., a static wire configuration that creates an elongated magnetic trapping potential. 
While such platforms have been successfully utilized in various experiments, many interesting future research applications such as quantum field thermal machines proposed by \cite{gluza_quantum_2021} require significantly more involved potential landscapes. 

A promising path that offers almost arbitrary flexibility is the use of optical dipole potentials (see \cite{grimm_optical_2000}) due to intense (laser) light. Spatial control of the incident laser beam for BECs using digital micro-mirror devices (DMDs) was shown in \cite{gauthier_direct_2016} and \cite{tajik_designing_2019}. Combined with a suitable optical imaging system, the two-dimensional array of mirrors of the DMD allows precise control of the intensity along the condensate's elongated direction. However, obtaining the required setting of the DMD to achieve a desired effective potential is highly non-trivial. 

While optimization-based control methods could be used in principle to obtain suitable binary patterns for the DMD offline using a detailed mathematical model of the system, the effects of a number of unknown disturbances limits their applicability in practice. Thus, heuristic algorithms to find suitable DMD settings from measurements of the atom densities were demonstrated in \cite{tajik_designing_2019}. Depending on the desired precision of the resulting potential, such algorithms require a significant number of iterations to converge. Since density measurements are destructive, i.e., one has to repeat the experiment consecutively, this heuristic approach is quite time consuming.

The application of more sophisticated control schemes that utilize model information to speed up convergence seems highly beneficial. Iterative learning control (ILC) methods (see, e.g., \cite{bristow_survey_2006}) have been applied to various optical problems such as pulse-shaping of short and ultra-short pulses in \cite{ren_laser_2011} and \cite{deutschmann_modeling_2018} or spatial shaping of deformable mirrors in \cite{cichy_modeling_2017}. 


Following these ideas, this paper introduces an ILC-based optical potential shaping algorithm for one-dimensional Bose-Einstein condensates by using measurements of the atomic density. To this end, we will shortly summarize the ingredients required to describe the system behavior mathematically in \prettyref{sec:model} and subsequently formulate the problem we intend to solve in \prettyref{sec:problem}. \prettyref{sec:ilc} first reformulates the potential shaping task as a tracking problem and an auxilliary mapping of binary DMD patterns onto quasi-continuous input values. While the latter is solved by using an offline optimization approach, the former is solved online using an ILC scheme based on density measurements. \prettyref{sec:simulation} finally illustrates the performance of the potential shaping algorithm by simulation.


%


\section{Mathematical model}
\label{sec:model}

\begin{figure}[b]
	\begin{center}
		\includegraphics[width=\columnwidth]{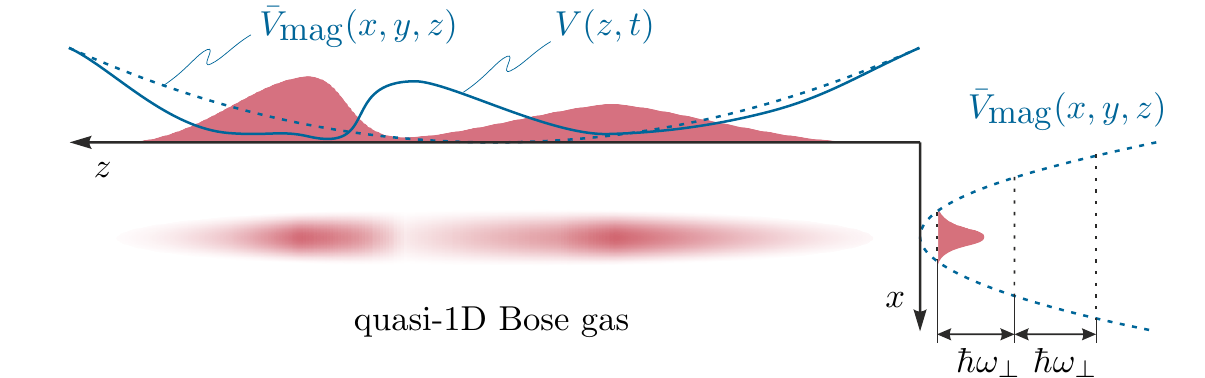}
		\caption{Density $\rho(t)$ of a cigar-shaped atomic cloud with tight transversal and weak longitudinal confinement due to the magnetic trapping potential $V_{\textrm{mag}}(x,y,z)$ of the atom chip. Using spatially shaped attractive or repulsive light fields to create an additional optical dipole potential $V_\textrm{opt}(z,t)$, one can generate complex longitudinal potentials $V(z,t)$.}
		\label{fig:atomchip}
	\end{center}
\end{figure}

We consider a Bose-Einstein condensate (BEC) in a highly elongated magnetic trap on an atom chip that yields a harmonic magnetic potential $\bar{V}_\textrm{mag}(x,y,z)$ with the Cartesian coordinates $x,y,z$. Along the weakly confined longitudinal direction $z \in \mathcal{D} = [-L/2,L/2]$ we can additionally generate a tunable optical dipole potential $V_\text{opt}(z,t)$ by using spatially shaped light fields. The dynamics of the resulting cigar-shaped quasi-1D condensate of $N$ particles can be effectively described by a non-polynomial Schrödinger equation (npSE) (see, e.g., \cite{salasnich_effective_2002}) for its longitudinal wave function $\Psi(z,t)$ in normalized form
\begin{align}\label{eq:NPSE}
	i \partial_{t} \Psi(z,t) = &- \frac{1}{2m} \partial_{zz} \Psi(z,t) + V(z,t)\Psi(z,t) \\
	&+ \omega_{\perp} \left(\frac{1+3a_{s} N |\Psi(z,t)|^2}{\sqrt{1+2a_{s} N |\Psi(z,t)|^2}} - 1\right)\Psi(z,t),\nonumber
\end{align}
with the (normalized) mass $m$, the transversal scattering length $a_s$ and the transversal angular frequency $\omega_{\perp}$ given by the (normalized) magnetic potential $\bar{V}_\textrm{mag}(x,y,z) = \frac{\omega_\perp}{2}  (x^2 + y^2) + V_\textrm{mag}(z)$. The effective longitudinal potential is thus given by $V(z,t) = V_\textrm{mag}(z) + V_\text{opt}(z,t)$. The probability density $\rho(z,t)$ of finding a particle of the atomic cloud at position $z$ and at time $t$ can be obtained by $\rho(z,t) = |\Psi(z,t)|^2$. Note that \prettyref{eq:NPSE} assumes that the condensate remains in its ground state in transversal direction, as illustrated in \prettyref{fig:atomchip}, which is the case for sufficiently low temperatures $T$, i.e., $k_{\textrm{B}} T < \hbar \omega_{\perp}$ with the Boltzmann constant $k_{\textrm{B}}$ and the reduced Planck constant $\hbar$.

\begin{rem}
	For simplicity, we assume that the length scales at the condensate and the DMD are identical, i.e., that the optical setup does not introduce any magnification, and that the length of the DMD aligns with the considered domain $\mathcal{D}$ of the condensate. We furthermore assume that the actual extension of the condensate as well as the incident beam is considerably smaller than $\mathcal{D}$, i.e., that the magnetic trapping potential is sufficiently large such that the density vanishes at its boundaries. This way, one can furthermore use the standard convolution notation on an infinite spatial domain, see \prettyref{eq:optical_system}.
\end{rem}

\begin{figure}[tbp]
	\begin{center}
		\includegraphics[width=\columnwidth]{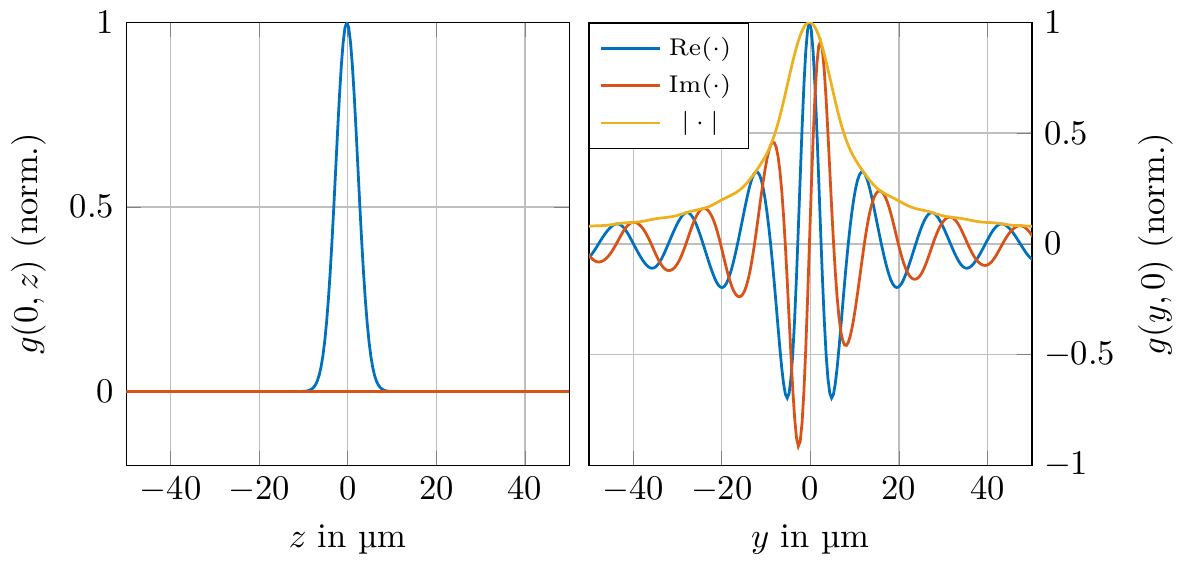}
		\caption{Spatial low-pass filtering due to rectangular apperture: while the longitudinal direction $g(0,z)$ is a simple Gaussian point-spread function, the transversal direction $g(y,0)$ is significantly broadened.}
		\label{fig:behavior_optics}
	\end{center}
\end{figure}

To create an adjustable optical potential, far-detuned coherent light with intensity $I(z,t)$ is directed at the condensate with a resulting proportional dipole potential $V_\text{opt}(z,t) = \alpha_V I(z,t)$ with $\alpha_V > 0$, e.g., see \cite{grimm_optical_2000}. An incident beam with transversal beam profile of the electric field $E_\textrm{in}(y,z)$ is directed at a DMD of pixel width $\Delta_\textrm{DMD}$. Assuming that the gap between individual pixels is negligible, the DMD can be described as a spatial reflectance $R(y,z,t)$ that is given by the binary input matrix $\vec{u}(t) \in \mathcal{U} = \mathbb{B}^{n_R \times n_C}$, with 
\begin{align}
	\label{eq:DMD_reflectance}
	R(y,z,t) = u_{ij}(t) 
\end{align}
if $(i - 1/2) \Delta_\textrm{DMD} \leq y < (i+1/2) \Delta_\textrm{DMD}$ and $(j -1/2)\Delta_\textrm{DMD} \leq z < (j+1/2) \Delta_\textrm{DMD}$ for $i = 1,\ldots, n_R$ and $j = 1,\ldots, n_C$. Feeding the reflected beam through an imaging system for filtering, the resulting output beam is in general given by the convolution (omitting time $t$ for brevity)
\begin{equation}
	\label{eq:optical_system}
	E_{\textrm{out}}(y,z) = \int_{-\infty}^{\infty} \int_{-\infty}^{\infty} g(y-\xi,z-\eta) E_{\textrm{in}}(\xi,\eta) R(\xi,\eta) \dd \xi \dd \eta
\end{equation}
with the point-spread function $g(y,z)$ of the optical stage. The intensity at the condensate is finally given by $I(z,t) = |E_{\textrm{out}}(0,z,t)|^2$ by placing the condensate in the plane $y=0$ without loss of generality, and thus the optical potential follows as
\begin{equation}
	\label{eq:optical_potential}
	V_\text{opt}(z,t) = \alpha_V |E_{\textrm{out}}(0,z,t)|^2.
\end{equation}
To expand the achievable dynamic range of the intensity at each position $z$ without limiting the available spatial resolution in longitudinal direction, spatial averaging in transversal direction is used by including a rectangular apperture in the Fourier plane of the imaging system, see \cite{tajik_designing_2019}. Specifically, the therein considered setup can be described by a point-spread function as given in \prettyref{fig:behavior_optics}. While the longitudinal direction $g(0,z)$ is approximated by a simple Gaussian point-spread function, the transversal direction $g(y,0)$ is significantly broadened and introduces coherent effects due to the relative phase shift that laser light obtains when reflected from different pixels.

\begin{figure}[tbp]
	\begin{center}
		\includegraphics[width=\columnwidth]{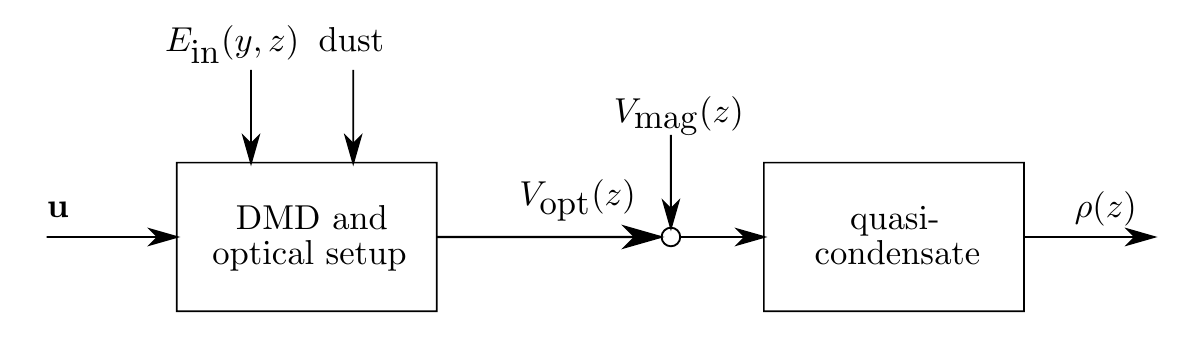}
		\caption{Graphical representation of the complete system. The (quasi-) condensate elongated in $z$ direction with density $\rho(z)$ is manipulated via an optical potential $V_{\textrm{opt}}(z)$. Its precise shape can be adjusted by manipulating the incoming light field $E_{\textrm{in}}(y,z)$ using the individual pixels of a DMD.}
		\label{fig:system}
	\end{center}
\end{figure}

While the transversal profile of the incident laser beam $E_\textrm{in}(y,z)$ is typically of Gaussian shape and aligned with the optical axis, i.e., $E_\textrm{in}(y,z) \approx |E_\textrm{in}| p_y(y)p_z(z)$ with $p_y(y) = \exp\left( - \frac{y^2}{\sigma_{in,y}^2}\right)$ and $p_z(z) = \exp\left(- \frac{z^2}{\sigma_{in,z}^2} \right)$, its detailed structure is typically unknown and may change slowly over time. Similarly, the longitudinal magnetic potential is only approximately quadratic, i.e., $V_\textrm{mag}(z) \approx \frac{\omega_\parallel}{2} z^2$ with $\omega_\parallel \ll \omega_\perp$. Finally, the optical setup may introduce deviations due to dust particles, imperfect aligment or spatial inhomogeneities of optical elements. These imperfections are considered as disturbances perturbing the desired operation.
An overview of the complete system is given in \prettyref{fig:system}, whereby the dependence on time was omitted for simplicity.

\section{Problem formulation}
\label{sec:problem}

Obtaining desired time-varying 1D potentials is of high interest for many application of BECs. However, applying potential shaping algorithms to dynamic settings is not straightforward, since one has to disentangle imperfections of the applied potentials from effects of the dynamics of the BEC due to \prettyref{eq:NPSE}.
We will therefore restrict ourselves to stationary scenarios for the remainder of this paper. 

For a given DMD pattern $\vec{u}^n \in \mathcal{U}$ during the $n$-th iteration, one obtains the resulting potential $V(z;\vec{u}^n) = V_\textrm{mag}(z) + V_\textrm{opt}(z;\vec{u}^n)$ from \prettyref{eq:optical_system} and \prettyref{eq:optical_potential}.
The groundstate $\varphi^n(z)$ in this potential is given by the nonlinear eigenvalue problem
\begin{align}
	\label{eq:eigenvalue_NLSE}
	\mu^n \varphi^n(z) = &- \frac{1}{2m} \partial_{zz} \varphi^n(z) +V(z;\vec{u}^n) \varphi^n(z)  \\
	            &+ \omega_{\perp} \left(\frac{1+3a_{s} N |\varphi^n(z)|^2}{\sqrt{1+2a_{s} N |\varphi^n(z)|^2}} - 1\right)\varphi^n(z) \nonumber,
\end{align}
with the corresponding smallest eigenvalue (or chemical potential) $\mu^n$ and the corresponding densitiy $\rho^n(z) = | \varphi^n(z) |^2$ measurable through absorption imaging techniques \cite{tajik_designing_2019}. 

We thus want to obtain an abstract learning operator $\mathcal{L}: \mathcal{U} \otimes L^2(\mathcal{D}) \mapsto \mathcal{U}$ such that the learning law
\begin{align}
	\label{eq:learning_law}
	\vec{u}^{n+1} = \mathcal{L} (\vec{u}^n , e_\rho^n(z)),
\end{align}
with some suitably defined measurement error $e_\rho^n(z)$, iteratively converges towards the desired potential $V^d(z)$ with the corresponding density $\rho^d(z)$ given by \prettyref{eq:eigenvalue_NLSE} irrespective of the initial potential $V^0(z)$.

\section{Iterative potential shaping with quasi-continuous input mapping}
\label{sec:ilc}

To obtain such a learning operator, ILC-based methods seem to be a natural choice. Two features of the problem at hand are quite problematic for existing ILC concepts: First, ILC-based methods are typically limited to reference tracking problems, i.e., where the output trajectory over a certain (temporal or spatial) domain shall track a desired reference by adjusting the input trajectory over the \emph{same} domain. Second, ILC strategies usually require continuous adjustability of the input quantities. 
We will thus reformulate the problem in such a way that we obtain a (quasi-) continuously adjustable tracking problem before applying estabilished ILC methods in the following sections.

\subsection{Reformulation into a tracking problem}

Since we assume that the optical setup (and in particular the rectangular apperture) is aligned with the condensate, the point-spread function of the optics can be decomposed into $g(y,z) = g_y(y) g_z(z)$.
Together with the approximately Gaussian input beam, one can thus separate the optical behavior \prettyref{eq:optical_system} along the condensate into their longitudinal and transversal parts, i.e.,
\begin{align*}
	&E_{\textrm{out}}(y,z) = \int_{-\infty}^{\infty} \int_{-\infty}^{\infty} g(y-\xi,z-\eta) E_{\textrm{in}}(\xi,\eta) R(\xi,\eta) \dd \xi \dd \eta \\
	&= \!\int_{-\infty}^{\infty} \!\!\! g_z(z-\eta) p_z(\eta) \underbrace{|E_{\textrm{in}}| \int_{-\infty}^{\infty} \!\!\! g_y(y-\xi)p_y(\xi) R(\xi,\eta) \dd \xi}_{=E^\perp (y,\eta)} \dd \eta.
\end{align*}
Using that $g_z(z)$ is Gaussian together with \prettyref{eq:optical_potential}, one obtains the approximate relation for the  optical potential
\begin{align}
	\label{eq:optical_potential_Eperp}
	V_{\textrm{opt}}(z) &\approx \alpha_V \left[\int_{-\infty}^{\infty} g_z(z-\eta) p_z(\eta) |E^\perp (0,\eta)| \dd \eta \right] ^2
\end{align}
For the purpose of control design, we give the following interpretation:
The longitudinal potential is mainly governed by superpositions due to the Gaussian point-spread function blurring the fictitous quantity $|E^\perp (0,z)|$ weighted by the beam inhomogeneity  $p_z(z)$, which is a linear process except for the quadratic action. The value of $E^\perp (0,z)$ at a position $z$ is given by the pattern of $R(y,z)$ in the transversal direction, i.e., the column of $\vec{u}$ associated with $z$ by \prettyref{eq:DMD_reflectance}. Manipulating the optical potential $V_\textrm{opt}(z)$ (and hence $\rho(z)$) due to $E^\perp(0,z)$ is therefore a standard tracking problem, whereby quasi-continuous values of $E^\perp(0,z)$ have to be generated using the respective column of $\vec{u}$.

\subsection{Quasi-continuous input mapping}
We thus search for transversal patterns of $R(y,z)$ through the respective column of $u_{ij}$ such that $|E^\perp(0,z)|$ equals a desired value at some arbitrary $z \in \mathcal{D}$. 
One easily sees that this task is not solvable in a unique way. One may additionally try to avoid strong variations of $E^\perp$ in transversal direction $y$ for robustness reasons and to reduce unwanted transversal gradients. Moreover, the achievable values are bounded by $E^\perp_\textrm{max}$ due to the available intensity of the incident beam and one can introduce the normalized quantity $\tilde{E}^\perp(y,z) = \frac{E^\perp(y,z)}{E^\perp_\textrm{max}} \in [0,1]$. Since the given problem is independent of the longitudinal position, i.e., $z$ or $i$, respectively, we will consider columns $\vec{u}^\perp \in \mathbb{B}^{n_C}$ of $\vec{u}$ only and use $\tilde{E}^\perp(y)$ instead of $\tilde{E}^\perp(y,z)$ for the following.

\begin{figure}[tbp]
	\begin{center}
		\includegraphics[width=\columnwidth]{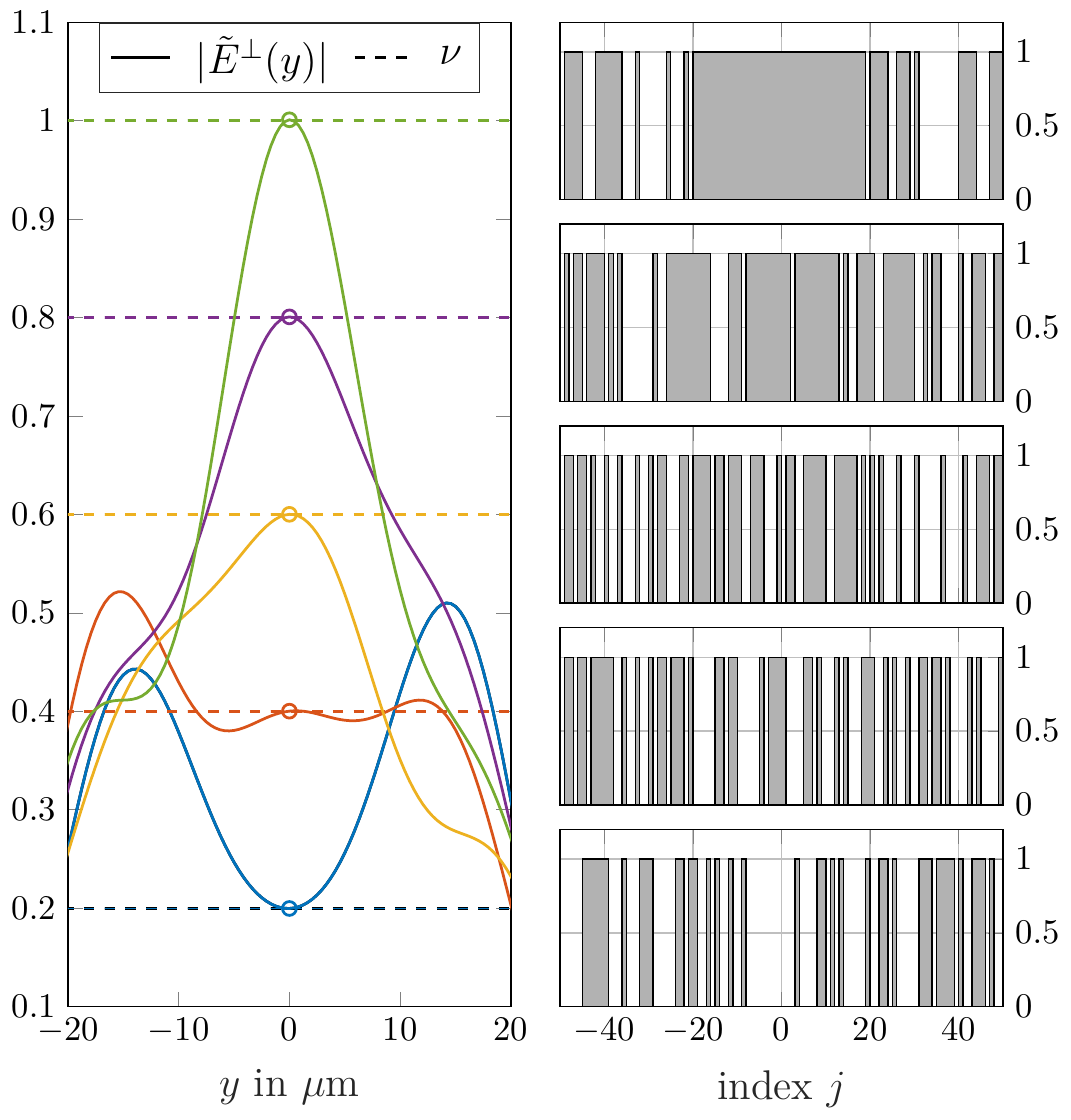}
		\caption{Exemplary solutions of \prettyref{eq:input_optimization_problem} illustrating the transversal behavior $|E^\perp(y)|$ (left) for the corresponding binary DMD pattern $u^\perp_j$ (right) for $\nu \in \{0.2, 0.4,0.6,0.8,1\}$, $n_\nu = 51$, $n_C = 100$, $\Delta y = 4 \Delta_\textrm{DMD}$, and $\gamma_\perp = 0.3$.}
		\label{fig:input_mapping}
	\end{center}
\end{figure}

\begin{figure}[tbp]
	\begin{center}
		\includegraphics[width=0.9\columnwidth]{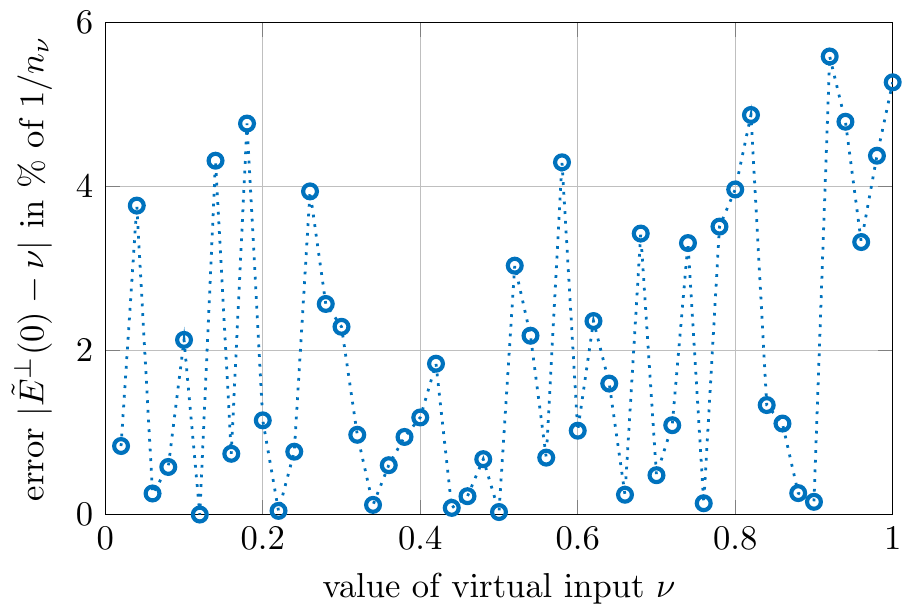}
		\caption{Error $|\tilde{E}^\perp(0) - \nu|$ due to the input mapping $\vec{f}(\nu)$ in percent of the step width $1 / n_\nu$ for $n_\nu = 51$ equally spaced values of $\nu \in [0,1]$.}
		\vspace{-0.5cm}
		\label{fig:mapping_error}
	\end{center}
\end{figure}

To adjust $\tilde{E}^\perp(0)$ to some desired value $\nu \in [0,1]$, one can find suitable transversal patterns $\vec{u}^\perp$ by solving the optimization problem
\begin{align}
	\label{eq:input_optimization_problem}
	\minimize{\quad (|\tilde{E}_\perp(0)| - \nu)^2 + \gamma_\perp \int_{-\Delta y}^{\Delta y}  (|\tilde{E}_\perp(\eta)| - \nu)^2 \dd \eta }{\vec{u}^\perp},
\end{align}
with $\gamma_\perp > 0$ to penalize deviations from $\nu$ in the region $y \in [-\Delta y, \Delta y]$.
Solutions of \prettyref{eq:input_optimization_problem} define a static mapping $\vec{f}^\perp(\nu) = \vec{u}^\perp$ that maps a virtual, quasi-continuous, and normalized input quantity $\nu$ onto desired transversal DMD patters $\vec{u}^\perp$. For practical implementation, one may simply choose to discretize the range of virtual input values into $n_\nu$ values and store the precomputed results in a look-up table.
Exemplary solutions\footnote{The binary values of $\vec{u}^\perp$ suggests the use of heuristic evolutionary algorithms. The results in this paper are based on the MATLAB 2020a genetic algorithm (ga) implementation.} are illustrated in \prettyref{fig:input_mapping} for $n_\nu = 51$, $n_C = 100$, $\Delta y = 4 \Delta_\textrm{DMD}$, and $\gamma_\perp = 0.3$. \prettyref{fig:mapping_error} shows the corresponding error $|\tilde{E}^\perp(0) - \nu|$ of the mapping $\vec{f}^\perp(\nu)$. Since the error remains within a few percent of the step width $1 / n_\nu$, the solution accuracy is sufficient such that $\vec{f}^\perp$ is monotonic.

Finally, $\vec{f}^\perp$ allows us to use $\nu(z)$ as a continuous proxy for $\vec{u}$ by defining the longitudinally discretized mapping $\vec{f}: L^2(\mathcal{D}) \mapsto \mathcal{U}$ according to
\begin{align*}
	\vec{f}(\nu(z)) = 
	\begin{bmatrix}
		\vec{f}^\perp(\nu(z_1)),\! & \vec{f}^\perp(\nu(z_2)), \!& \ldots , \!&\! \vec{f}^\perp(\nu(z_{n_R})),
	\end{bmatrix}
\end{align*}
with the (center) position of the DMD pixels $z_i = i \Delta_\textrm{DMD}$. Notice that this dicretization process does not introduce a significant error since the pixel size $\Delta_\textrm{DMD}$ is typically much smaller than the width of the Gaussian point-spread function of the imaging system.
Assuming that the input mapping is exact, i.e., $E^\perp(0,z) = E^\perp_\textrm{max} \nu(z)$,  the resulting optical potential \prettyref{eq:optical_potential_Eperp} is given by
\begin{align}
	\label{eq:optical_potential_nu}
	V_{\textrm{opt}}(z;\vec{u}) &\approx \alpha_V  \left[ E^\perp_\textrm{max} \int_{-\infty}^{\infty} g_z(z-\eta) p_z(\eta) \nu(\eta) \dd \eta \right]^2
\end{align}
where $\vec{u} = \vec{f}(\nu(z))$.
 
\subsection{Potential shaping using ILC methods}
\begin{figure}[htbp]
	\begin{center}
		\includegraphics[width=\columnwidth]{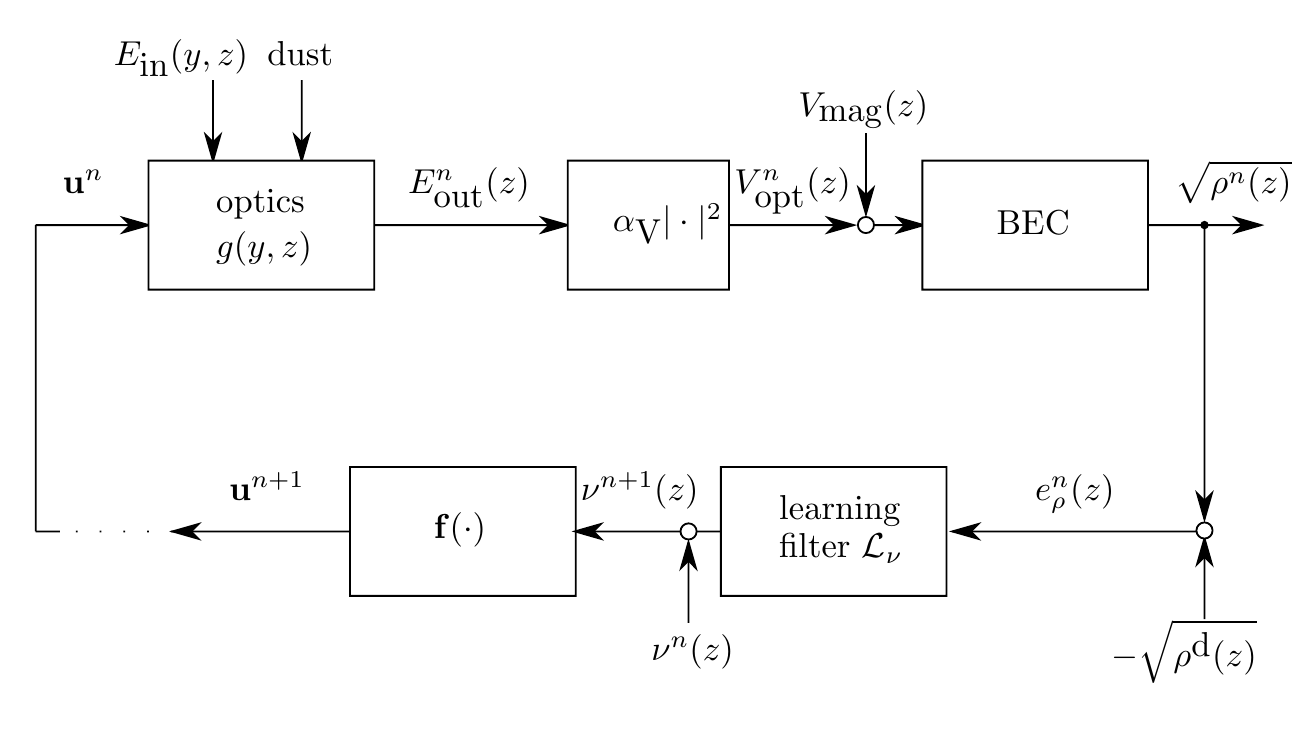}
		\caption{Schematic illustration of the iterative potential shaping using ILC methods: A standard ILC update is applied to the virtual input $\nu^{n+1}(z)$ due to the measured density $\rho^n(z)$. A corresponding DMD pattern $\vec{u}$ is subsequently generated using the input mapping $\vec{f}$, which assigns each column the associated pattern due to the optimization problem \prettyref{eq:input_optimization_problem}.}
		\label{fig:ILC}
	\end{center}
\end{figure}

Due to the reformulation using a (quasi-)continuous virtual input, the problem of shaping desired potentials can be treated using standard ILC methods.
Relating measured densities $\rho^n(z)$ of the $n$-th iteration with the corresponding virtual input $\nu^n(z)$ according to \prettyref{eq:eigenvalue_NLSE} is quite difficult in general. Neglecting the kinectic energy of the condensate (i.e., using $\partial_{zz} \varphi^n(z) \approx 0$), one directly obtains the so-called Thomas-Fermi-approximation  
\begin{align}
	\label{eq:TF_approx}
	V(z;\vec{u}^n) = \mu^n - \omega_{\perp} \left(\frac{1+3a_{s} N \rho^n(z)}{\sqrt{1+2a_{s} N \rho^n(z)}} - 1\right),
\end{align}
wherefore $\rho^n(z)$ is given by a static but nonlinear mapping of $V(z;\vec{u}^n)$. 
\begin{rem}
	Note that $\mu^n$ changes for different $\nu^n(z)$, but since this chemical potential is constant w.r.t. $z$, its value does not change the resulting density $\rho^n(z)$.
\end{rem}
Since the potential is only determined up to an additive constant, \prettyref{eq:TF_approx} has to be used with care. Introducing the measured density error as $e_\rho^n = \sqrt{\rho^n(z)} - \sqrt{\rho^d(z)}$, where $\rho^n(z)$ and $\rho^d(z)$ correspond to inputs $\vec{u}^n = \vec{f}(\nu^n)$ and $\vec{u}^d = \vec{f}(\nu^d)$, respectively, and assuming that $2a_{s} N \rho^n(z) < 1$, which is typically the case for the considered scenarios, one obtains 
\begin{align*}
	e_\rho^n &= \kappa \sqrt{\mu^n - V_\textrm{mag}(z) - V_\textrm{opt}(z;\vec{u}^n) }  \\ 
	&- \kappa \sqrt{\mu^d - V_\textrm{mag}(z) - V_\textrm{opt}(z;\vec{u}^d) }
\end{align*}
with $\kappa = 1/\sqrt{3 \omega_{\perp} a_s N}$. Using \prettyref{eq:optical_potential_nu} with $\nu^n(z) = \nu^d(z) + \Delta \nu^n(z)$, assuming that $\mu^n \approx \mu^d$ and expanding the first square root around $\Delta \nu^n(z) = 0$ finally yields the approximate description
\begin{align}
	\label{eq:density_nu}
	e_\rho^n &\approx - \alpha_\rho(z)  \int_{-\infty}^{\infty} g_z(z-\eta) p_z(\eta) \Delta \nu^n(\eta) \dd \eta
\end{align}
with $\alpha_\rho(z) = \kappa E^\perp_\textrm{max} \sqrt{\frac{\alpha_V (V^d(z) - V_\textrm{mag}(z)) }{\mu^d - V^d(z))}}$.
For the ILC design, we further simplify \prettyref{eq:density_nu} by replacing the spatial scaling $\alpha_\rho(z)$ with $\overbar{\alpha_\rho} = \max_{z\in \mathcal{D}} \alpha_\rho(z)$ and neglecting the inhomogeneous longitudinal beam shape, i.e., $p_z(\eta) = 1$. Thus, the system exhibits an approximately linear and spatially-invariant input-output behavior when considering the square-root of the density as output quantity.
%
For a given DMD setting $\vec{u}^n = \vec{f}(\nu^n(z))$, one can feed the measured error $e_\rho^n(z)$ into a standard learning filter $\mathcal{L}_\nu: L^2(\mathcal{D}) \mapsto L^2(\mathcal{D})$ to obtain the updated virtual input according to
\begin{align}
	\label{eq:transversal_learning_law}
	\nu^{n+1}(z) &= \nu^n(z) + \mathcal{L}_\nu e^n_\rho(z) \\ 
	             &= \int_{-\infty}^{\infty} L_\nu(z-\xi) e_\rho^n(\xi) \dd \xi \nonumber,
\end{align}
with the (convolutional) learning kernel $L_\nu(z)$. The updated DMD setting follows from $\vec{u}^{n+1} = \vec{f}(\nu^{n+1}(z))$ as illustrated in \prettyref{fig:ILC}. 
The abstract learning operator $\mathcal{L}$ in \prettyref{eq:learning_law} is thus formally given by $\mathcal{L}(\vec{u} , e_\rho(z) ) = \vec{f}\left(  \vec{f}^{-1}(\vec{u}) + \mathcal{L}_\nu  e_\rho(z)  \right)$.

\begin{figure}[htbp]
	\begin{center}
		\includegraphics[width=0.9\columnwidth]{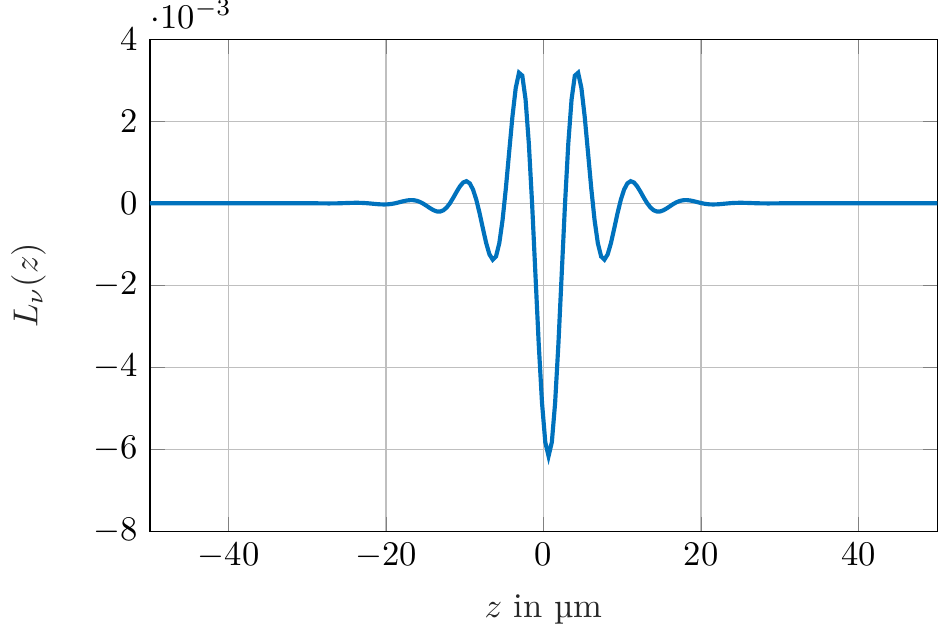}
		\caption{Learning kernel $L_\nu(z)$ according to \prettyref{eq:kernel} for $\gamma_\nu = \num{1e-2} \max  |G(jk)|^2$.}
		\label{fig:kernel}
	\end{center}
\end{figure}

To find suitable learning kernels $L_\nu(z)$, the LTI nature of the approximate system description \prettyref{eq:density_nu} suggests the use of frequency-domain ILC methods, e.g., see \cite{boeren_frequency-domain_2016}, which can directly incorporate the infinite-dimensional dynamics of \prettyref{eq:density_nu}, see \cite{deutschmann-olek_stochastic_2021}. Using the spatial Fourier transform $f(jk) = \mathcal{F}\left\{ f(z) \right\} = \int_{-\infty}^{\infty} f(z) \exp(-j k z) \dd k$, the transfer function of the system is given by
\begin{align}
	G(jk) = -\mathcal{F}\left\{ \overbar{\alpha_\rho} g_z(z)\right\}
\end{align}
and a simple pseudo-inverse learning filter (cp. \cite{ghosh_pseudoinverse-based_2002}) in the frequency-domain reads as
\begin{align}
	\label{eq:kernel}
	L_\nu(jk) = \frac{G^*(jk)}{\gamma_\nu + G^*(jk) G(jk)}
\end{align}
using the regularization parameter $\gamma_\nu > 0$ and $(\cdot)^*$ denoting the complex conjugate. 
The resulting learning kernel $L_\nu(z)$ is illustrated in \prettyref{fig:kernel} for $\gamma_\nu = \num{1e-2} \max  |G(jk)|^2$. 
Notice that the resulting learning filter is - due to the assumptions above - equivalent to established noncausal MMSE (Wiener) reconstruction filters for white noise that are commonly used in imaging applications, e.g., \cite{oppenheim_signals_2016,gonzalez_digital_2003}.

\section{Simulation scenarios}
\label{sec:simulation}

In the following, we want to illustrate the proposed control strategy with a simulation scenario based on the specifications given in \cite{tajik_designing_2019}. For this, the optical apparatus is simulated using \prettyref{eq:optical_system} with the point-spread function illustrated in \prettyref{fig:behavior_optics} with a Gaussian profile of width $\sigma = 2.5 \si{\micro\metre}$ and a transversal filtering due to an apperture opening of \SI{1}{\milli\metre} in the Fourier plane of the optics. The used profile of the incident beam $E_\textrm{in}(y,z)$ was measured experimentally and is roughly Gaussian in shape with $\sigma_{in,y} = \SI{13}{\micro\metre}$ and $\sigma_{in,z} = \SI{125}{\micro\metre}$, respectively, while the amplitude $|E_\textrm{in}|$ (i.e., the power of the laser) was chosen such that the desired potential can be achieved with the available input range. The magnetic trapping potential is essentially quadratic with $\omega_\parallel = 2\pi \times\SI{7}{\hertz}$ and additionally includes sinusoidal deviations with spatial wavelength of $\SI{10}{\micro\metre}$, which is within the achievable resolution of the optical potentials due to the finite bandwidth of the optics. For the condensate described by \prettyref{eq:NPSE}, the normalized mass $m = \SI{1.368}{\milli\second\per\micro\metre\squared}$ (which corresponds to Rb87 atoms), the transversal frequency $\omega_{\perp} = 2\pi \times \SI{1.4}{\kilo\hertz}$ and the scattering length $a_s = \SI{5.2}{\nano\metre}$ for $N = 5000$ atoms are used. 
While the solution of the nonlinear eigenvalue problems is quite involved in general, the groundstate solution for the nonlinear Schrödinger equation \prettyref{eq:eigenvalue_NLSE} can be obtained from standard numerical methods using so-called imaginary time evolution techniques, see \cite{chiofalo_ground_2000} for details.

For the quasi-continuous input mapping, a look-up table with $n_\nu = 51$ values as illustrated in \prettyref{fig:input_mapping} is used together with the learning kernel $L_\nu$ in \prettyref{fig:kernel} due to \prettyref{eq:kernel} for the transversal learning law \prettyref{eq:transversal_learning_law}. The initial virtual input is chosen as $\nu^0(z) = 0.5$ for equal control margins in both directions.
Finally, the desired potential is chosen as
\begin{align}
	\label{eq:desired_potential}
	V^d(z) = 
	\begin{cases}
		\frac{V_\text{max}}{2} \left( 1 + \cos(k_V z)\right) \quad &\text{for } |z| \leq \pi / k_V\\
		V_\text{max} \quad & \text{else},\\
	\end{cases}
\end{align}
with $V_\textrm{max} = 2\pi \times \SI{8}{\kilo\hertz}$ and $k_V = \SI{7.53e-2}{\per\micro\metre}$, which yields a harmonic double well potential with $\SI{41.8}{\micro\metre}$ width of each well.

\begin{figure}[htbp]
	\begin{center}
		\includegraphics[width=\columnwidth]{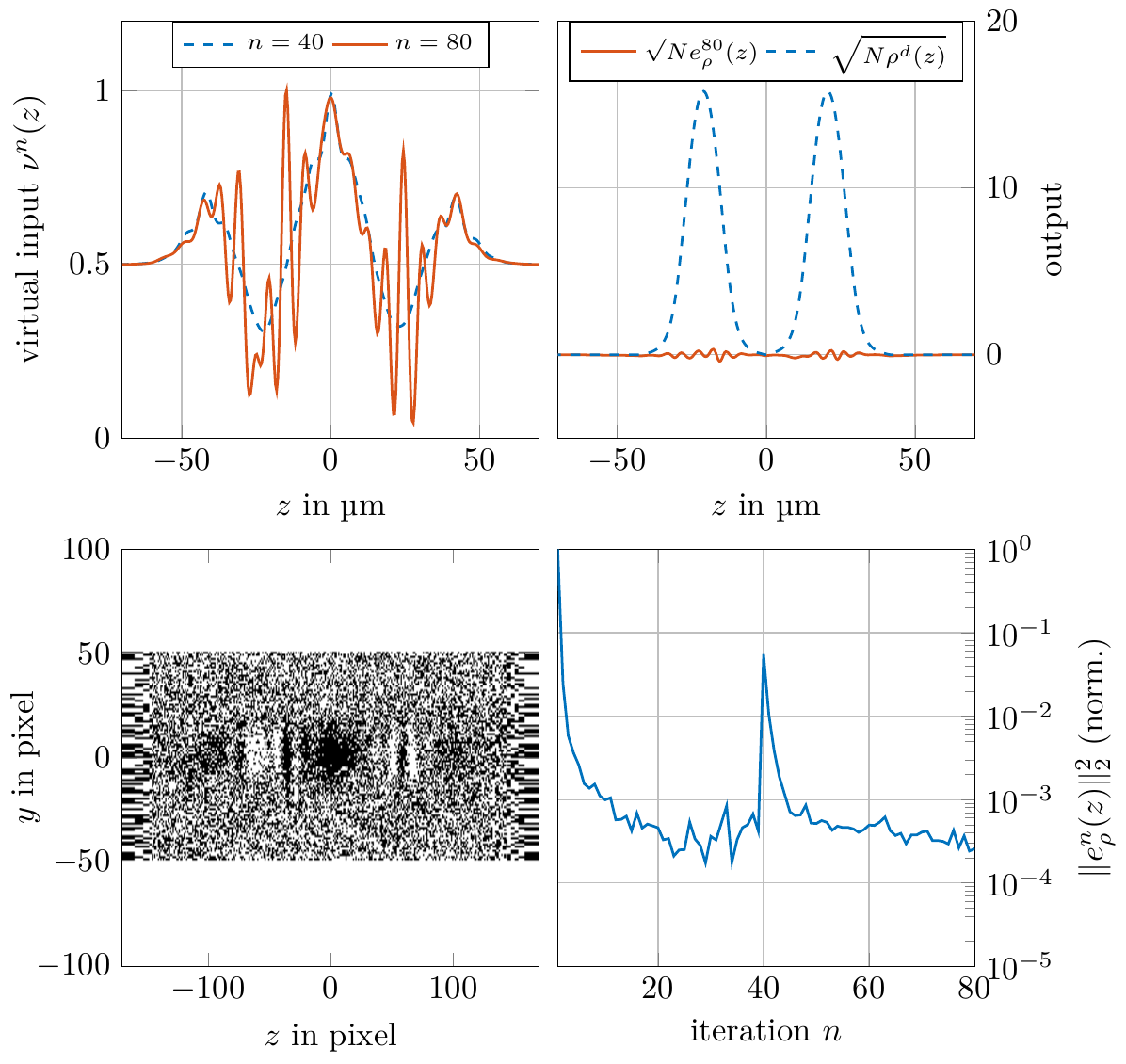}
		\caption{Simulation of the proposed potential shaping approach for the desired double-well potential \prettyref{eq:desired_potential} with added dark spots for iteration $n \geq 40$. The converged virtual input with ($\nu^{80}(z)$) and without ($\nu^{40}(z)$) dark spots of the optics (top left) is shown together with the corresponding DMD input pattern $\vec{u}^{80}$ (bottom left) and the resulting measured density error $e^{80}_\rho(z)$ for an expected desired density $\rho^d(z)$ (top right). The $L^2$-norm of the density error converges rapidly from the initial state as well as after the disturbances occur (bottom right).}
		\label{fig:density_iterations}
	\end{center}
\end{figure}

\prettyref{fig:density_iterations} illustrates the behavior of the closed-loop system. On the bottom right picture, one can see that the learning law reduces the density error by a factor of 100 during the first two iterations and progressively removes the remaining errors afterwards. Further simulation scenarios suggest that the remaining error for $n = 20, \ldots, 39$ is mainly determined by the limited resolution of the quasi-continuous input mapping and errors due to disturbances beyond the optical bandwidth. The virtual input $\nu^{40}(z)$ is given in the top left plot and shows the expected modulation required to generate the desired sinusoidal potential with additional corrections for the imperfections due to the magnetic trapping potential $V_\textrm{mag}$. At $n = 40$, additional disturbances in the form of small spots with lower transmissivity of the output lenses are added. These disturbances are again corrected quickly by the learning law, but their localized nature (partially beyond the bandwidth of the optical system) introduces quite severe corrective actions into the final virtual input $\nu^{80}(z)$. The corresponding DMD pattern is given in the lower left plot of \prettyref{fig:density_iterations}, whereby the pixel size was chosen to match the longitudinal spacing of the virtual input above.

\begin{figure}[htbp]
	\begin{center}
		\includegraphics[width=\columnwidth]{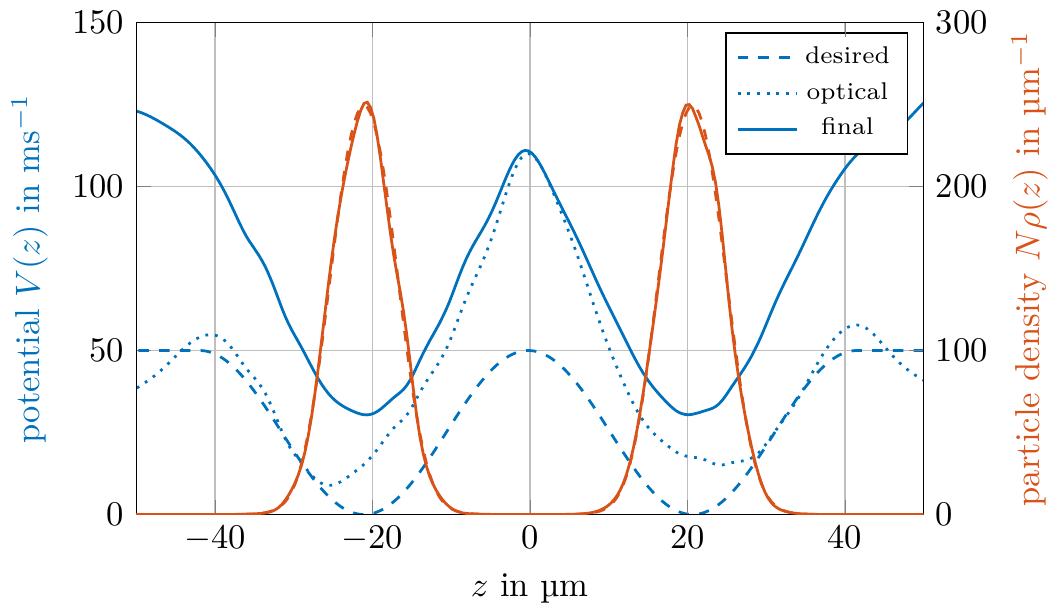}
		\caption{Final shape of the effective potential $V^{80}(z)$ (solid), its optical contribution (dotted) and the desired potential $V^d(z)$ (dashed) compared to the desired and measured densities.}
		\label{fig:density_potential}
	\end{center}
\end{figure}

The resulting final potential shapes are shown in \prettyref{fig:density_potential}. Since the magnetic potential rapidly increases away from the center position, the required contribution of the optics decreases. Notice that the vertical offset of the final and the desired potential is arbitrary since it is without physical meaning. The final shape of the density agrees quite well with the desired density. However, one can clearly see that regions of the potential where $\rho(z) \approx 0$ are not corrected by the learning law. This is expected since the particles do not explore these regions in steady state and thus any deviations of the effective potential remain hidden.

\bibliography{bibliography} 
\bibliographystyle{IEEEtran}

%
%
%
%

\end{document}